\documentclass{article}
\usepackage{amsmath,amssymb,graphicx,cite,times}

\begin{document}

\begin{center}
{\large\bf POLARIZATION OF TAU LEPTONS PRODUCED IN
           QUASIELASTIC NEUTRINO--NUCLEON SCATTERING}
\vskip 4mm
{\bf Konstantin\,S.\,Kuzmin},$^{1,2}$
{\bf Vladimir\,V.\,Lyubushkin}$^{1,3}$
and 
{\bf Vadim\,A.\,Naumov}$^{1,4}$
\vskip 3mm
{\small
$^{(1)}$ {\small\em Joint Institute for Nuclear Research, Dubna, Russia}
\\
$^{(2)}$ {\small\em Institute for Theoretical and Experimental Physics,
         Moscow, Russia}
\\
$^{(3)}$ {\small\em Physics Department of Irkutsk State University,
         Irkutsk, Russia}
\\
$^{(4)}$ {\small\em Dipartimento di Fisica and INFN, Sezione di Firenze,
         Sesto Fiorentino, Italy}
}
\end{center}

\vskip 4mm

\begin{abstract}
A numerical analysis of the polarization vector of $\tau$'s
produced through quasielastic ${\nu_\tau}$ and $\overline{\nu}_{\tau}$
interactions with nucleons is given with two models for vector
electromagnetic form factors of proton and neutron.
The impact of $G$ parity violating axial and vector second-class
currents is investigated by applying a simple heuristic model for
the induced scalar and tensor form factors.

\end{abstract}


\section{Introduction}
\label{sec:Introduction}

The success of many current and future experimental projects for
exploring neutrino oscillations and decay, nonstandard neutrino
interactions, 
baryon number nonconservation and related phenomena revolves
around the unambiguous reconstruction of $\tau$ lepton events
generated in neutrino-matter interactions and detected through
the secondary particles produced in the $\tau$ decays.
For instance, the $\tau$ lepton events provide a clear signature
of $\nu_\mu-\nu_\tau$ mixing; besides they are a source
of unavoidable background to the proton decay and $n\overline{n}$
transition experiments.
The momentum configuration of the $\tau$ decay secondaries is
a functional of the $\tau$ lepton polarization; the latter is
therefore a substantial input parameter for the data processing
in the mentioned experiments.

In recent paper by Hagiwara {\it et al.}~\cite{Hagiwara:03} the exact
formulas for the polarization density matrix of a lepton produced in
${\nu}N$ and $\overline{\nu}N$ interactions have been derived within
a noncovariant approach and under the Standard Model assumptions.
This result has been generalized in our work~\cite{Kuzmin:03} within
a covariant method and by taking into account the nonstandard
contributions induced by vector and axial second-class currents
(SCC)~\cite{LlewellynSmith:72,Grenacs:85,Wilkinson:00-01,Gardner:01}.
Since the Standard Model contains the first-class currents (FCC) only,
it is instructive to examine possible manifestations of the SCC induced
interactions and grasp the uncertainties due to SCC in the future
accelerator and astrophysical neutrino experiments.
In this letter, by applying the relevant results of Ref.~\cite{Kuzmin:03},
we briefly discuss some consequences of the SCC induced contributions
potentially measurable in future experiments on quasielastic (QE)
$\nu_\tau$ and $\overline{\nu}_{\tau}$ interactions with nucleons. 

\section{FCC effects}
\label{sec:FFE}

The hadronic currents describing the $\Delta Y=0$ processes
${\nu}_{\tau}n\to\tau^-p$ and $\overline{\nu}_{\tau}p\to\tau^+n$
may be written as  \cite{LlewellynSmith:72}
\begin{equation*}
\langle p,\,p'|\widehat{J}_\alpha^+|n,\,p\rangle=
\cos\theta_C\overline{u}_p\left(p'\right)\varGamma_\alpha\,u_n(p),
\quad
\langle n,\,p'|\widehat{J}_\alpha^-|p,\,p\rangle=
\cos\theta_C\overline{u}_n\left(p'\right)\overline{\varGamma}_\alpha\,u_p(p),
\end{equation*}
where $\theta_C$ is the Cabibbo angle and the vertex function
\begin{equation*}
\varGamma_\alpha=\gamma_\alpha F_V
              +i\sigma_{\alpha\beta}\frac{q_\beta}{2M}F_M
              + \frac{q_\alpha}{M} F_S
              + \left(\gamma_\alpha F_A
              + \frac{p_\alpha+p'_\alpha}{M} F_T
              + \frac{q_\alpha}{M}F_P\right)\gamma_5.
\end{equation*}
is defined through the six, in general complex, form factors
$F_i\left(q^2\right)$: FCC induced ($i=V,M,A,P$) and
SCC induced ($i=S,T$).
Here $p$ and $p'$ are, respectively, the 4-momenta of the initial
and final nucleons, $q=p'-p$ is the 4-momentum transfer and $M$
is the nucleon mass (as usually, the half-sum of proton and neutron
masses). Hereafter we follow the notation of Ref.~\cite{Kuzmin:03}.

We investigate two models for the Sachs form factors of proton
and neutron: the ``benchmark'' dipole model~\cite{LlewellynSmith:72}
with the standard value of vector mass ($M_V=0.84$\,GeV$/c^2$) and
the Gari--Kr\"uempelmann (GK) model~\cite{Gari:92} extended
and fine-tuned by Lomon~\cite{Lomon:02} to match current experimental
data. Specifically we explore the so-called ``GKex(02S)'' model which
fits the modern and consistent older data well and meets the requirements
of dispersion relations and of QCD at low and high 4-momentum transfer.

Note that the GKex(02S) model is very close numerically the ``BBA-2003''
parametrization by Budd {\it et al.}~\cite{Budd:03} obtained through a global
fit to the data from Rosenbluth analysis of elastic $ep$ cross section
measurements and those from the polarization transfer techniques.
Both these models do not take into account the two-photon contribution
corrections to the form factors (see Ref.~\cite{Arrington:04} and
references therein) as well as the most recent data from JLab and MAMI
(which are however in good agreement with the GKex(02S) and BBA-2003).

For the axial and pseudoscalar form factors we use the conventional
parametrizations
\begin{equation*}\label{F_AP}
F_A\left(q^2\right)=F_A(0)\left(1-\frac{q^2}{M^2_A}\right)^{-2}
\quad\text{and}\quad
F_P\left(q^2\right)=\frac{2M^2}{m^2_\pi-q^2}F_A\left(q^2\right)
\end{equation*}
with $F_A(0)=g_A=-1.2695\pm0.0029$.
The presently available experimental data on the axial mass, $M_A$,
show very wide spread, from roughly $0.6$ to 1.2~GeV/$c^2$
\cite{Bernard:01} with the weighted average of $1.026\pm0.021$\,GeV$/c^2$.
Side by side with the pseudoscalar form factor, $F_P\left(q^2\right)$,
this is the main source of uncertainties for predicting the ${\nu_\tau}n$
and $\overline{\nu}_{\tau}p$ QE cross sections and polarization effects.
Since the pseudoscalar form factor contribution enters multiplied by
$m_\ell/M$, it is small for the electron and muon production but
substantial for $\tau$ production, especially for small $\left|q^2\right|$
(since $\left|F_P\right|\gg\left|F_A\right|$ for $\left|q^2\right| \ll M^2$).
Let us remark that the ``standard'' expression for the pseudoscalar form
factor is only a (doubtful) parametrization inspired by the PCAC
hypothesis and the assumption that the pion pole dominates at
$\left|q^2\right|\lesssim m_\pi^2$ \cite{LlewellynSmith:72}.
A discussion of some other phenomenological models for the $F_P$ and
their impact to the $\tau$ lepton production cross section and polarization
is given recently by Hagiwara {\it et al.}~\cite{Hagiwara:04}.
For that reason we do not concern this important issue in present study.

\begin{figure}[htb]
\includegraphics[width=\linewidth]{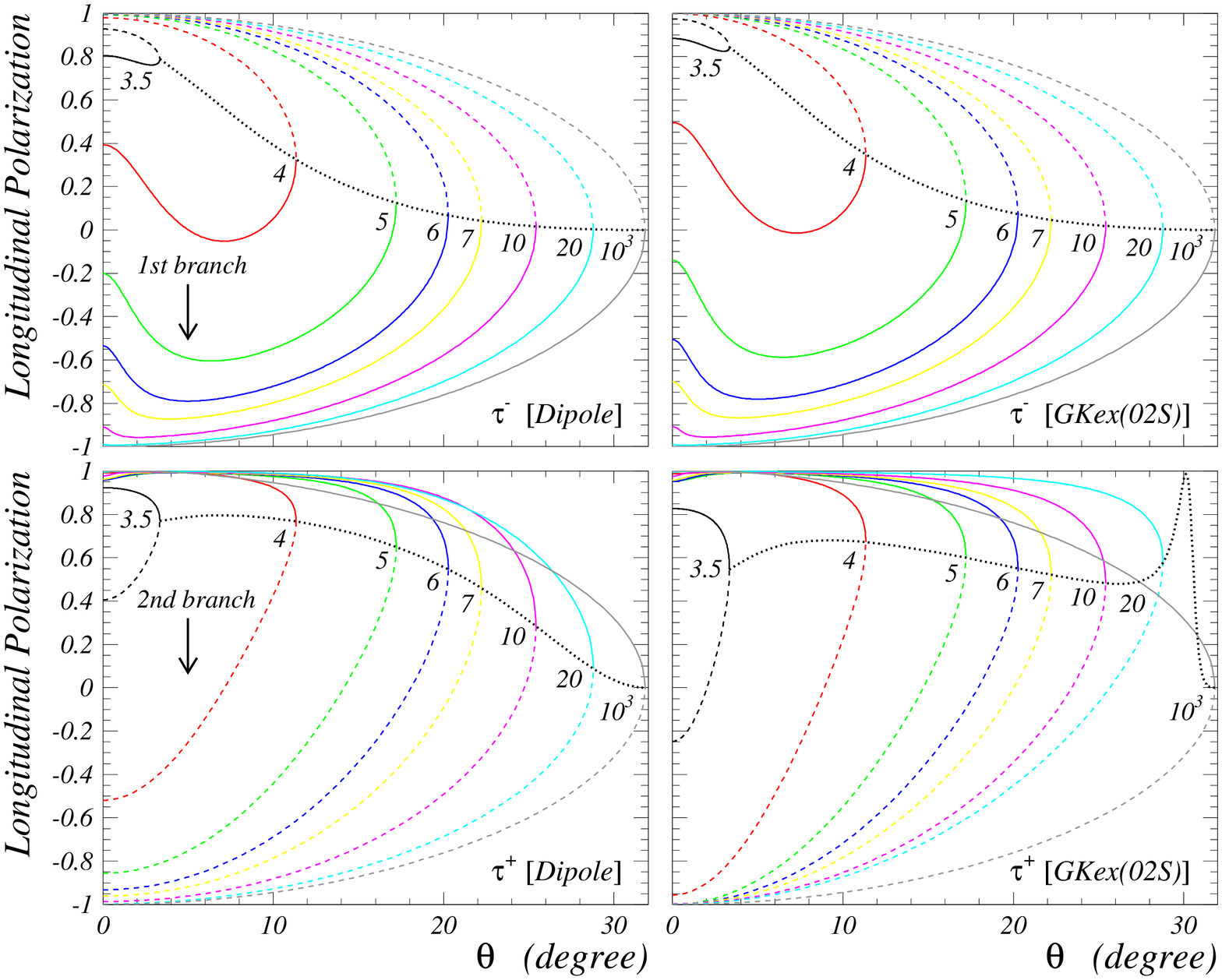}
\caption{Longitudinal polarization, ${\mathcal P}_L$,
         evaluated with the dipole and GKex(02S) models for
         the nucleon electromagnetic form factors at
         different (anti)neutrino energies (shown near the
         curves in GeV). The gray dotted curves indicate
         the boundaries between the two kinematically
         allowed solutions. The main kinematic branches are
         shown by solid curves.
\label{fig1}}
\end{figure}

\begin{figure}[htb]
\includegraphics[width=\linewidth]{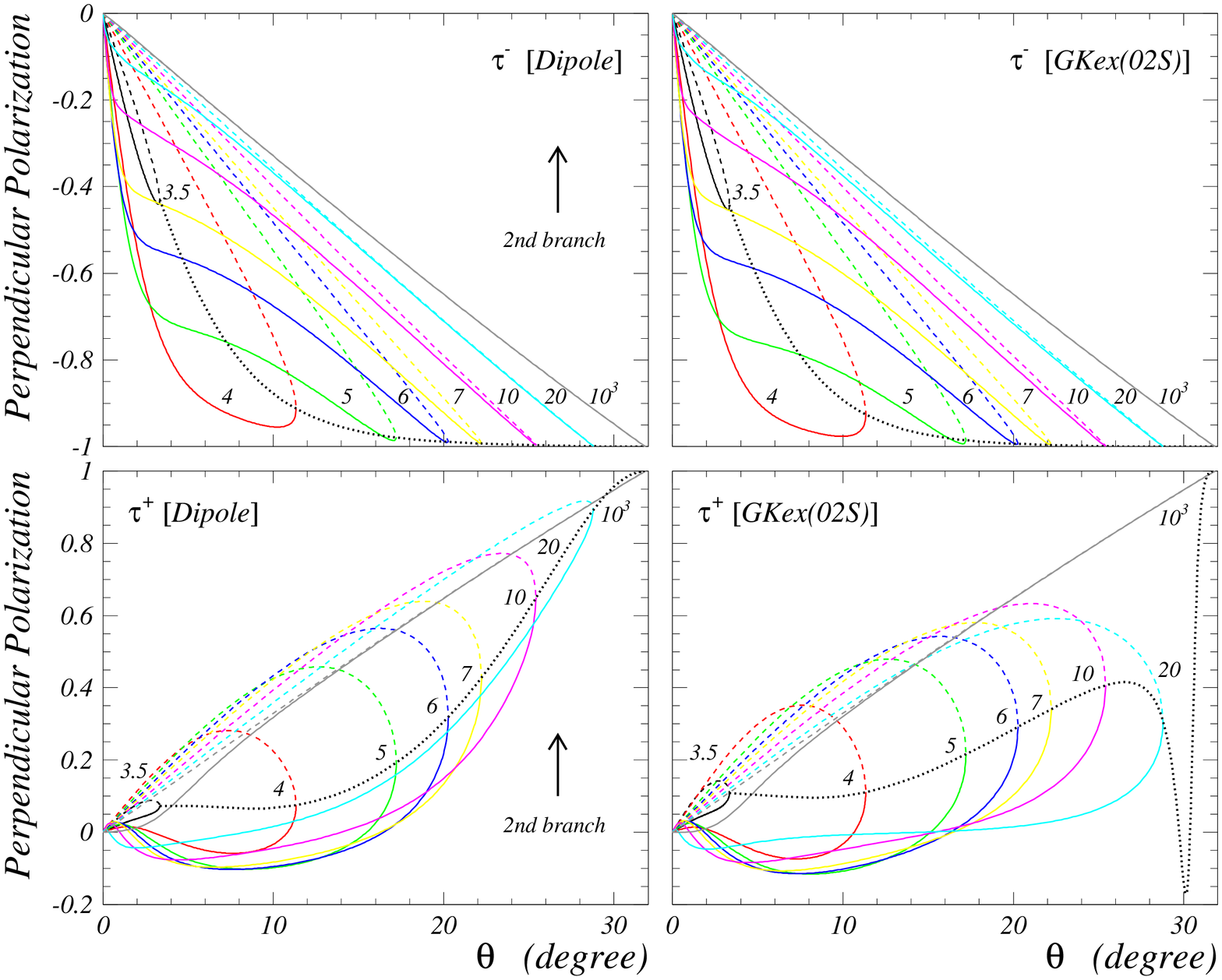}
\caption{Perpendicular polarization, ${\mathcal P}_P$,
         evaluated with the dipole and GKex(02S) models for
         the nucleon electromagnetic form factors at
         different (anti)neutrino energies (shown near the
         curves in GeV). The gray dotted curves indicate
         the boundaries between the two kinematically
         allowed solutions. The main kinematic branches are
         shown by solid curves.
\label{fig2}}
\end{figure}

\begin{figure}[htb]
\centering
\includegraphics[height=0.3724\linewidth]{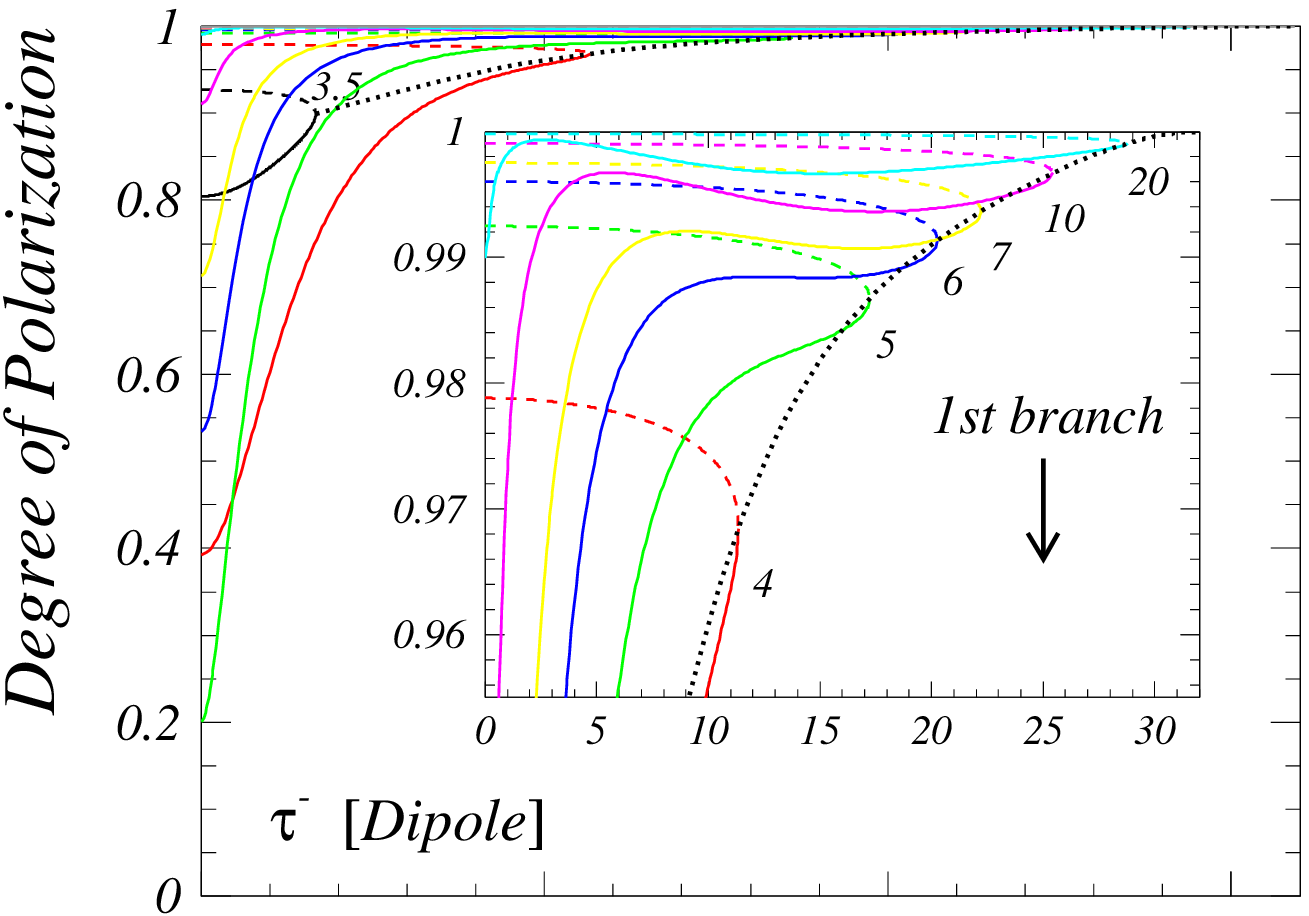}
\includegraphics[height=0.3724\linewidth]{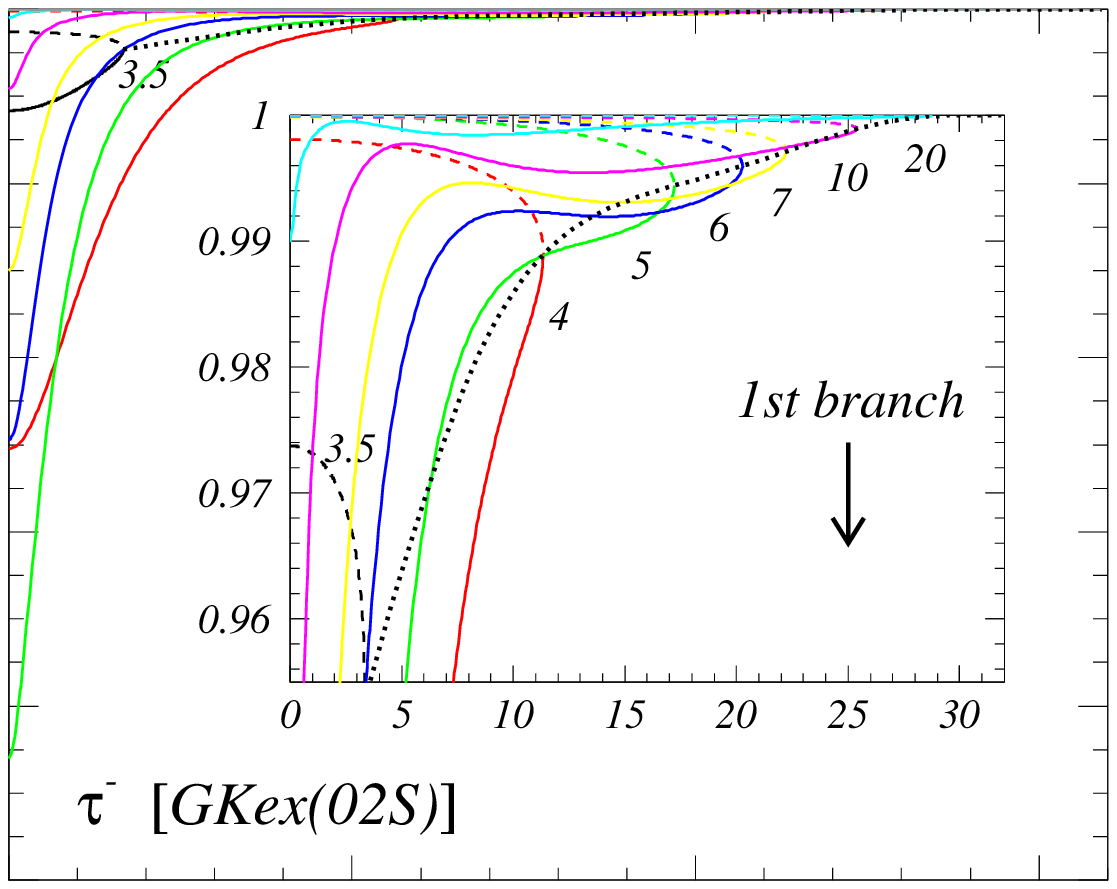}
\includegraphics[width=       \linewidth]{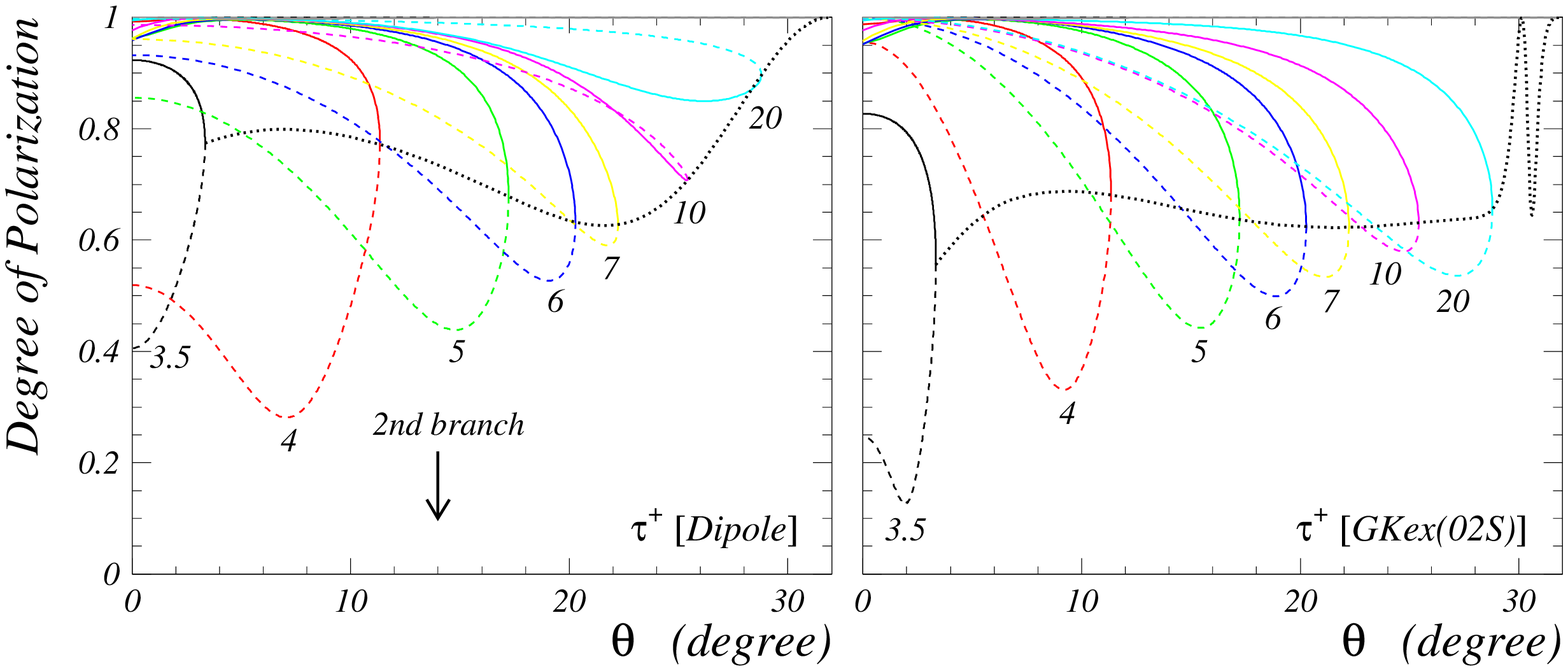}
\protect\caption{Degree of polarization,
                 $\left|\boldsymbol{\mathcal{P}}\right|$,
                 evaluated with the dipole and GKex(02S) models for
                 the nucleon electromagnetic form factors at different
                 (anti)neutrino energies (shown near the curves in GeV).
                 The gray dotted curves indicate the boundaries between
                 the two kinematically allowed solutions. The main kinematic
                 branches are shown by solid curves. The insets in two upper
                 panels show a zoomed view. 
\label{fig3}}
\end{figure}

For evaluating the lepton polarization vector
$\boldsymbol{\mathcal{P}}=
\left({\mathcal P}_P,{\mathcal P}_T,{\mathcal P}_L\right)$
we apply generic formulas given in Ref.~\cite{Kuzmin:03}.
In this section we will neglect the nonstandard contributions.
The transversal polarization, ${\mathcal P}_T$, is therefore exactly zero.
The longitudinal polarization, ${\mathcal P}_L$, perpendicular polarization,
${\mathcal P}_P$, and Lorentz invariant degree of polarization
\[
\left|\boldsymbol{\mathcal{P}}\right|=
\sqrt{{\mathcal P}_L^2+{\mathcal P}_P^2+{\mathcal P}_T^2}
\]
evaluated with the dipole and GKex(02S) models for the vector
electromagnetic form factors and with $M_A=1$\,GeV$/c^2$ are shown
in figs.~\ref{fig1}, \ref{fig2} and \ref{fig3}, respectively, as
functions of the lepton scattering angle $\theta$ at several
(anti)neutrino energies (both variables are measured in the laboratory
frame). Due to the large $\tau$ lepton mass, $m_\tau$, for each
allowed pair of variables ($E_\nu,\theta$) there are two solutions
(``branches'') of the energy-momentum conservation equations with
respect to the lepton momentum value $P_\tau=\left|\mathbf{p}_\tau\right|$.
These are given by
\begin{equation*}
P_{\tau}^\pm(\theta)
=\frac{E_\nu^*\left(ME_{\tau}^*\cos\theta\pm
  m_{\tau}E_N^*\sqrt{\zeta^2-\sin^2\theta}\right)}
       {M^2+\left(E_\nu^*\right)^2\sin^2\theta},
\qquad
\zeta=\frac{\left(E_\nu^*+E_N^*\right)P_{\tau}^*}{m_{\tau}E_\nu}.
\end{equation*}
Here $P_\tau^*$ is the lepton momentum, $E_\nu^*$, $E_\tau^*$ and
$E_N^*$ are the energies of (anti)neutrino, lepton and target
nucleon, respectively, measured in the center-of-mass frame.
The physically allowed interval of scattering angles is
\[
0\le\theta\le\arcsin(\zeta)\le\arcsin\left(M/m_\tau\right)
\approx 31.9^\circ.
\] 
The main kinematic branch, $P_{\tau}^+(\theta)$, whose contribution
into the total QE cross section is dominant, is shown in the figures
by solid curves and the second branch, $P_{\tau}^-(\theta)$, is
shown by dashed curves.
The dotted curves display the boundary between the two branches
which is defined by the condition $\sin\theta=\zeta$
[and thus $P_\tau=2m_\tau^2E_\nu\cos\theta/\left(2ME_\nu+m_\tau^2\right)$].

First of all, as is seen from the figures, the behavior of the $\tau$
lepton polarization is quite different from the naively expected.
At low energies and small scattering angles (the most important
kinematic range for underground neutrino experiments), the degree
of polarization of $\tau^-$ ($\tau^+$) is far from unity (close to
unity) for the main branch. There always exist the nontrivial
perpendicular (to the reaction plane) polarization for both $\tau^-$
and $\tau^+$ but their dependencies from $E_\nu$ and $\theta$ are
distinctly different for  $\tau^-$ and $\tau^+$.
It is also clear that the components of the polarization vector as well as
the degree of polarization of $\tau^-$ are almost insensitive to the model
(except for the near-threshold energies). This is in clear contrast to the
case of $\tau^+$ lepton, for which the polarization vector is very sensitive
to the shapes of the vector electromagnetic form factors for essentially
any antineutrino energy.

\section{SCC effects}
\label{sec:SCC}

A major problem in the quantitative analysis of the SCC effects is
the $q^2$ dependence of the induced scalar and tensor form factors.
To get some feeling for how big the SCC effects could be, let us consider
the following \emph{toy} model for the form factors:
\begin{align}
\label{F_S}
F_S\left(q^2\right)&=
\xi_Se^{i\phi_S}F_V(0)\left(1-\frac{q^2}{M^2_S}\right)^{-2}, \\
\label{F_T}
F_T\left(q^2\right)&=
\xi_Te^{i\phi_T}F_A(0)\left(1-\frac{q^2}{M^2_T}\right)^{-2}.
\end{align}
The model includes six free parameters, $\xi_{S,T}\ge0$, $\phi_{S,T}$
and $M_{S,T}$ and is an elementary amplification of the models adopted
by experimental collaboration at
FNAL~\cite{Bell:78}, IHEP (Serpukhov)~\cite{Belikov:85} and
BNL~\cite{Baker:81,Ahrens:88} to constrain the SCC strengths from
the measurements of $\nu/\overline{\nu}$ interactions with nucleons.
All the measurements~\cite{Bell:78,Baker:81,Belikov:85,Ahrens:88} found
no significant indication of nonzero SCC contributions.

The strongest 90\% C.L. upper limit on the axial SCC strength $\xi_T$
has been obtained at the Brookhaven Alternating Gradient Synchrotron
(BNL AGS) experiment with a wide-band $\overline{\nu}_{\mu}$
beam with $\langle E_{\overline{\nu}}\rangle=$ 1.2 GeV, $|q^2|<1.2$
(GeV$/c$)$^2$ \cite{Ahrens:88} as a function of the ``tensor mass''
$M_T$, assuming CVC ($\xi_S=0$) 
and the dipole model for the Sachs form factors of the proton.
The limit ranges from about 0.78 at $M_T=0.5$ GeV$/c^2$ to about 0.11 at
$M_T=1.5$ GeV$/c^2$. The BNL AGS constraint at the lower limit of
$M_T$ is not too informative since it is about 7 times weaker than
that follows from the nuclear structure studies~\cite{Wilkinson:00-01}
while at the upper limit it is appropriate for our heuristic exploration.

The BNL AGS constraint to the vector SCC (violating the standard CVC
hypothesis) is not so robust ($\xi_S<1.8$ at $M_S=1.0$ GeV/$c^2$ and
assuming $\xi_T=0$) and should not be taken too literally since the
scalar form factor contribution to the QE muon production cross
section is reduced by factor of $\left(m_\mu/M\right)^2\sim0.01$.
However this is the only, to our knowledge, experimental limit to
the CVC violation obtained at comparatively large momentum transfer.
The value of $M_A$ simultaneously obtained in the BNL AGS experiment
was 1.09 GeV/$c^2$ which is \emph{the same} as the value obtained
assuming no SCC.

Notice that there is no any reduction of the vector SCC contribution
to the QE $\tau$ production [$\left(m_\tau/M\right)^2\approx3.6$] and
hence even small violation of the CVC principle could affect both
the $\tau$ lepton production cross section and polarization.

\begin{figure}[hbt]
\includegraphics[width=\linewidth]{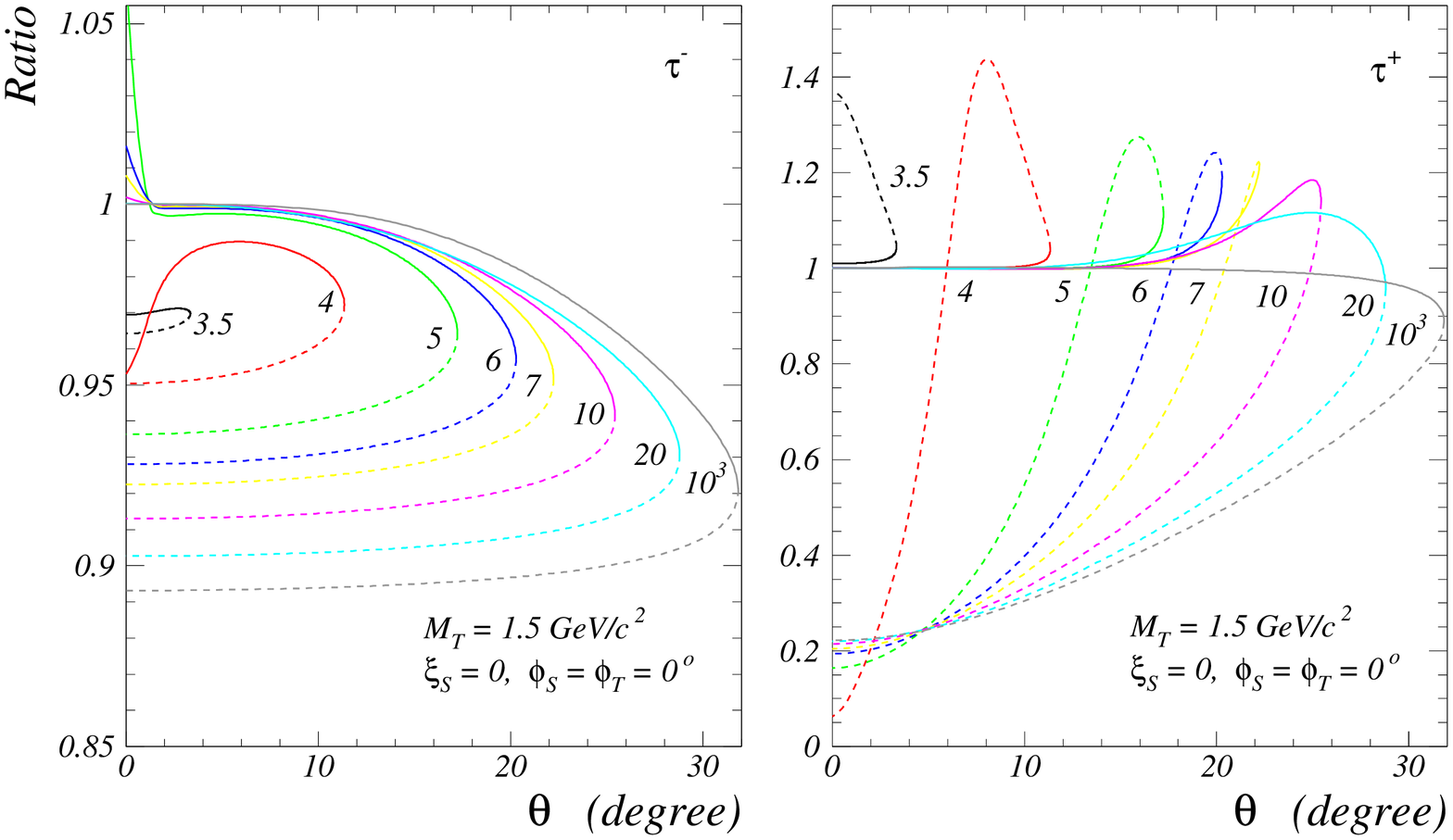}
\caption{Ratios  $\left|\boldsymbol{\mathcal{P}}\left(\xi_T=0.1\right)\right|/
         \left|\boldsymbol{\mathcal{P}}\left(\xi_T=0\right)\right|$
         evaluated with the dipole model for the nucleon
         electromagnetic form factors and with
         $M_A=1.0$ GeV/$c^2$. Values of the SCC parameters
         are given by the legends; (anti)neutrino
         energies are shown near the curves in GeV. The main
         kinematic branches are shown by solid curves.
\label{fig4}}
\end{figure}

\begin{figure}[htb]
\includegraphics[width=\linewidth]{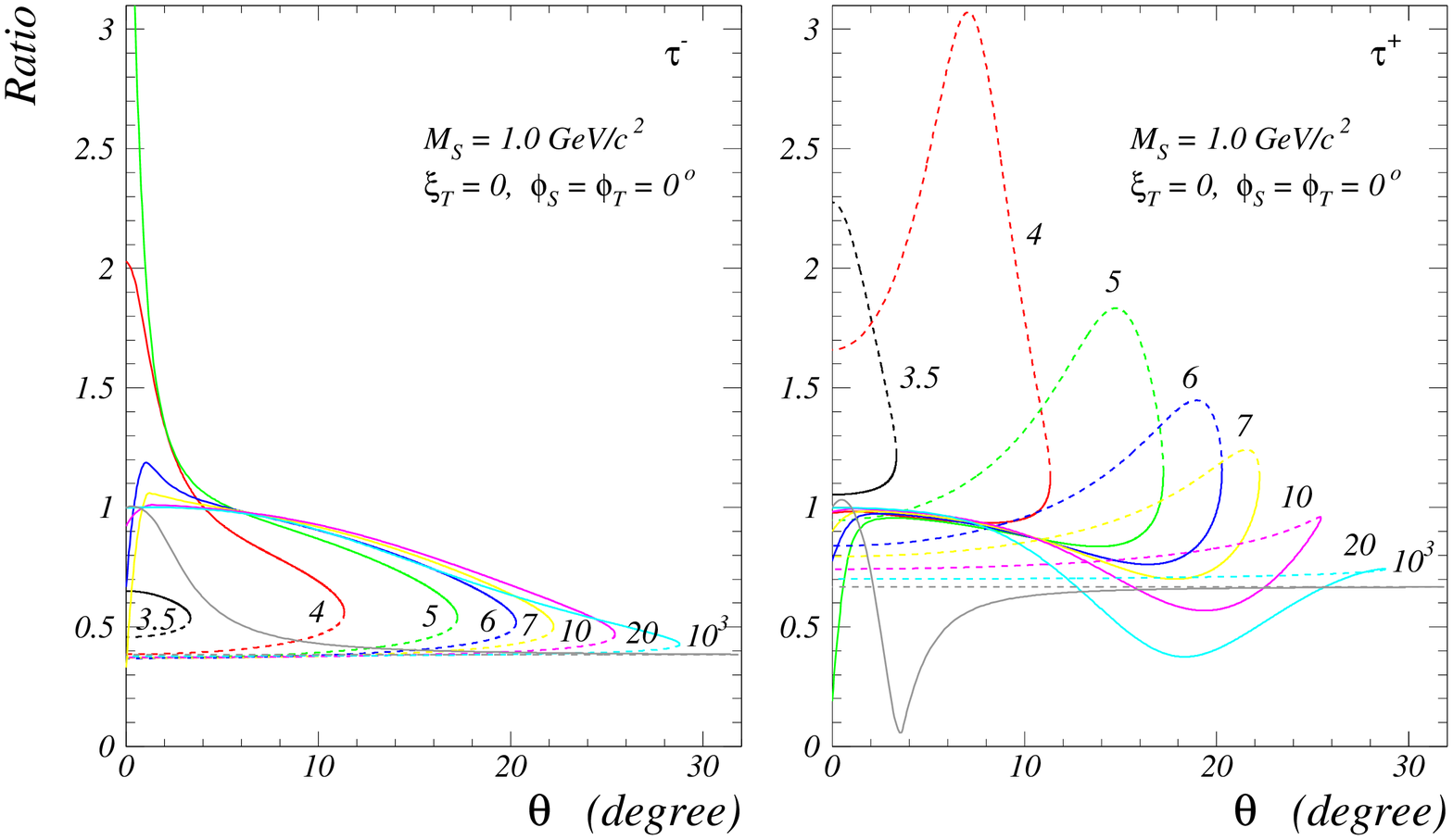}
\caption{Ratios $\left|\boldsymbol{\mathcal{P}}\left(\xi_S=1.8\right)\right|/
         \left|\boldsymbol{\mathcal{P}}\left(\xi_S=0\right)\right|$
         evaluated with the dipole model for the nucleon
         electromagnetic form factors and with
         $M_A=1.0$ GeV/$c^2$. Values of the SCC parameters
         are given by the legends; (anti)neutrino
         energies are shown near the curves in GeV. The main
         kinematic branches are shown by solid curves.
\label{fig5}}
\end{figure}

In figs.~\ref{fig4} and \ref{fig5} we show a representative example for
the ratios of the $\tau^\mp$ lepton degree of polarization,
$\left|\boldsymbol{\mathcal{P}}\right|$, evaluated with nonzero $\xi_{T,S}$
to those with $\xi_{T,S}=0$. Only the dipole model for the Sachs form factors
is utilized here since the relative SCC effects are qualitatively the same
for the GKex(02S) model. The calculations are done with $M_A=1.0$ GeV$/c^2$
and $\phi_{T,S}=0$. The latter restriction is equivalent to the time
reversal invariance while the nonzero strength parameters $\xi_{T,S}$
violate the charge symmetry of weak interactions.

Our choice of the unknown parameters $M_T$ and $\xi_T$ is made to ensure
the BNL AGS limits~\cite{Ahrens:88} (very sensitive to $M_T$) as well as
the restrictions on the axial SCC coupling constant from studies of $\beta$
decay of complex nuclei~\cite{Wilkinson:00-01} (much less sensitive to $M_T$).
As a conservative upper limit, we accept $\xi_T=0.1$ and $M_T=1.5$ GeV$/c^2$.
The values $M_S=1.0$ GeV$/c^2$ and $\xi_S=1.8$ are taken according to the
BNL AGS upper limit in order to demonstrate the maximum possible effect
of the vector current nonconservation.

As is seen from the figures, the impact of the tensor and (especially)
scalar form factors is not negligible in the kinematic regions where the
differential QE cross section $d\sigma/d\left|q^2\right|$ is comparatively
large (the main kinematic branch, small scattering angles). On the other
hand, the axial SCC contribution (which is not forbidden by any basic
principle like CVC) is not too dramatic and, especially for $\tau^+$,
the polarization vector is more sensitive to small variations of the
{\it standard} axial and pseudoscalar form factors~\cite{Hagiwara:04}. 
This is a handicap for experimental probe of the SCC effects by
${\nu_\tau}$ and $\overline{\nu}_{\tau}$. On the other hand, this is
an advantage for the current and future neutrino oscillation experiments
since the SCC contributions are not the main source of systematic
uncertainties.

Let us emphasize that similar effects for the muon polarization vector
are uninterestingly small; so we do not discuss the corresponding
numerical examples here.

\begin{figure}[htb]
\includegraphics[width=\linewidth]{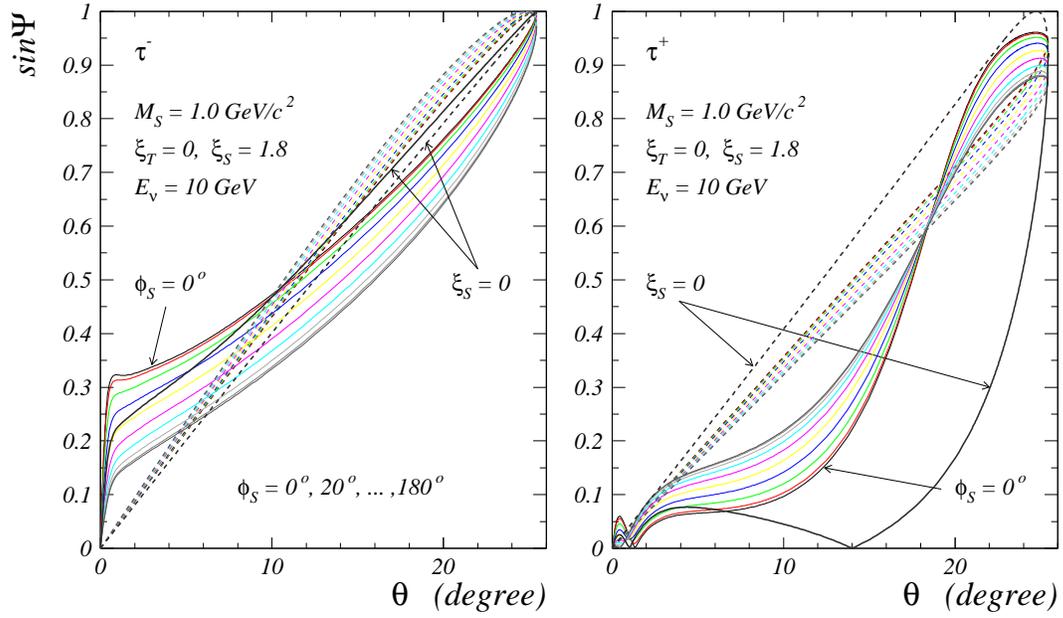}
\caption{$\sin\Psi$ as a function of scattering angle $\theta$ at
         (anti)neutrino energy $E_\nu=10$ GeV for different values of the
         ``scalar phase'' $\phi_S$ (from $0^\circ$ to $180^\circ$ with step
         of $20^\circ$).
         The calculations are done with the dipole model for the nucleon
         electromagnetic form factors  and with $M_A=1.0$ GeV/$c^2$.
         Values of the remaining SCC parameters are given by the legends.
         The main kinematic branches are shown by solid curves.
         The case $\xi_S=0$ is also  shown by thick gray curves.
\label{fig6}}
\end{figure}

For the last and more speculative possibility, we consider effects of the
phase factors $\exp\left(i\phi_{S,T}\right)$ violating the $T$ invariance.
It can be shown that, for the allowed values of the strength parameter
$\xi_T$, the lepton polarization vector is not significantly affected by
the ``tensor phase'' $\phi_T$. However it is not in general the case for
the ``scalar phase'' $\phi_S$. The effect is shown in fig.~\ref{fig6} for
the sine of $\Psi$, the angle between the momentum and polarization vector
of a $\tau^-$ ($\tau^+$) lepton generated quasielastically by a 10\,GeV
$\nu_\tau$ ($\overline{\nu}_\tau$). The angle $\Psi$ is defined by 
\[
\sin\Psi = \frac{\sqrt{{\mathcal P}_P^2+
           {\mathcal P}_T^2}}{\left|\boldsymbol{\mathcal P}\right|}
\quad\text{or}\quad
\cos\Psi = \frac{{\mathcal P}_L}
           {\left|\boldsymbol{\mathcal P}\right|}.
\]
Due to the nontrivial phase $\phi_S$, the transverse polarization,
${\mathcal P}_T$, is no longer exactly zero but remains comparatively
small. One can prove that $\left|{\mathcal P}_L\right|$,
$\left|{\mathcal P}_P\right|$ and $\left|{\mathcal P}_T\right|$ are
invariant under the transformation $\phi_S\mapsto2\pi-\phi_S$.
It is therefore sufficient to vary the scalar phase within the
range $0^\circ$ to $180^\circ$.
All the parameters used for this calculation are listed in the
legends of fig.~\ref{fig6}. The value $\xi_S=1.8$ is again taken
to show the maximum possible effect allowed by the BNL AGS limit.
As one can see from the figure, the vector SCC induced $C$ violating
effects may be rather large. Figure~\ref{fig6} also demonstrates
quite sizable transformation of $\sin\Psi$ when the phase $\phi_S$
varies from $0^\circ$ to $180^\circ$. 
For the main kinematic branch (solid lines), the $\sin\Psi$ is
a decreasing function of the phase $\phi_S$ for $\tau^-$ at
almost any scattering angle.
It is not the case for $\tau^+$: the $\sin\Psi$ is an increasing
function of $\phi_S$ at $2^\circ\lesssim\theta\lesssim18^\circ$
and rather nonmonotonic function for other allowed ranges of the
scattering angle.

We recall in conclusion that our analysis is only valid within the
adopted {\it ad hoc} model for the SCC induced form factors,
including the quite arbitrary choice for the tensor and scalar
mass values.

\section*{Acknowledgments}

We thank S.~M.~Bilenky, A.~V.~Efremov, B.~Z.~Kopeliovich,
D.~V.~Naumov, J.~Soffer and O.~V.~Teryaev for useful conversations.
V.~L. and K.~K. are grateful to the Physics Department of
Florence University for warm hospitality and financial
support of this work.

\clearpage

\begin{center}
{\bf\large FIGURES}
\end{center}

\end{document}